# 3D System Design: A Case for Building Customized Modular Systems in 3D

Philip Emma, Eren Kursun

IBM Research, T.J. Watson Research Center,
Yorktown Heights, NY, 10598
email: pemma@us.ibm.com

**Abstract**

3D promises a new dimension in composing systems by aggregating chips. Literally. While the most common uses are still tightly connected with its early forms as a packaging technology, new application domains have been emerging. As the underlying technology continues to evolve, the unique leverages of 3D have become increasingly appealing to a larger range of applications: from embedded/mobile applications to servers and memory systems. In this paper we focus on the system-level implications of 3D technology, trying to differentiate the unique advantages that it provides to different market segments and applications.

## 1. Introduction

In most cases used commercially today, 3D is very simply a packaging technique that is used to simplify integration. So far, its principle applications have been in high-volume markets where the costs of assembly are paramount (such as cameras and cell phones); and/or in markets where the physical size of the end-product is fixed (e.g., DIMMs); as well as in markets where both the power density and the inter-chip signal density are low. The goals of these practical uses are not to use technology in different ways; they are to make the end products simpler.

Besides raw density, 3D allows new degrees of freedom; none of them useful in *all* market sectors, but each useful in a subset (Figure 1). Whether they represent a real opportunity depends on exactly what you are trying to achieve, and your reason for doing trying to do so:

- By integrating multiple components into a stack, 3D enables a single package to suffice, and it simplifies the subsequent assembly processes to make the end product. It provides a way to continue the density scaling for a given footprint.
- The modular integration of layers can enable a range of products to be made from a common set of subsystems in the form of layers or chiplets. This has the effect of volumizing those subsystems, which reduces their costs and their times-to-market. 3D stacking can also be used to incorporate pieces of the electrical and service infrastructures directly, which can provide improved electrical performance. In addition, 3D can allow clocking, power delivery and control, and test-related logic to be incorporated in a more modular way.

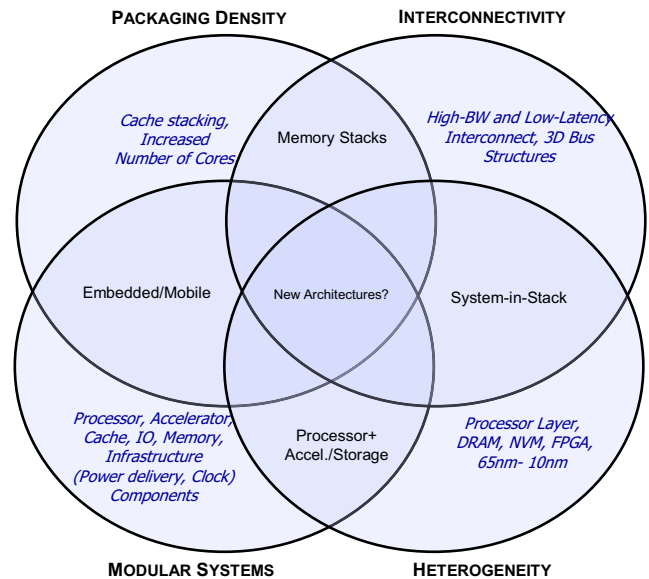

Figure 1. Basic dimensions in 3D system design

- 3D allows disparate technologies to be combined within a single stacked component in a manner that does not compromise either technology. This leverages emerging technologies being used together with traditional silicon technology.
- Scaling through-silicon-via (TSV) size and pitch in 3D enables high bandwidth and low latency interconnects among multiple device layers. This can enable massive internal bandwidth, which could be used to improve cache hierarchies, and to make faster and more efficient bus and interconnect structures.

Furthermore, the heterogeneity enabled by 3D can also be leveraged to incorporate various reconfigurability features, as well as specialized accelerators and I/O technologies for performance improvement in traditional systems. The potential benefits of 3D at the system-level will likely be dependent on the success of putting together customized systems out of modular sub-system components, and in effectively navigating the design space with potential advantages and disadvantages.

## 2. System-level Implications: The Basics

### 2. 1. Potential Advantages

From a system design perspective, one of the interesting questions is whether there are opportunities to make other aspects of systems (perhaps new kinds of systems) in 3D that are impractical in 2D, and whether those aspects require new technological innovations [2].

While the principle goal of consumer electronics is cost effectiveness, it is worth asking whether 3D could possibly enable new kinds of systems and new kinds of data processing that don't seem particularly economical (even fathomable) today, but that may be possible in the future. The first two generic opportunities (articulated earlier) are rather straightforward: enabling a simpler package to hold multiple chips, allowing modularity to create a range of products from a small number of parts, and allowing disparate technologies to be integral to a single component. The benefits here are obvious, and none particularly stresses the limits of 3D.

The third leverage adumbrates a more elegant electrical infrastructure, since different componentry can be integrated so as to yield cleaner power, and a more optimized service infrastructure for the kinds of testing, clocking, etc., that make the system more robust. Again, these enable direct physical improvements (which manifest as "simplifiers") to a system.

Last but not least in the advantages list, we suggest a system application in which the logical elements of a system can be physically co-located in the *{x,y}* dimensions so that unprecedented bandwidth in the *{z}* dimension can allow the stack to do types of computation that would not be fathomable in 2-space. With advances in the size, pitch and reliability of TSVs, it may be more likely to take advantages of this additional level of interconnectivity [6]. There have been considerable changes in the TSV sizes and pitches in the past years, and recent studies with TSV sizes of only a few microns indicate that this trend is continuing.

*2.1.a. Interconnectivity*
Having additional interconnectivity has been a topic of interest in a wide range of circuit and system-level studies. At a finer granularity, having increased interconnectivity with reduced latency has been seen as a vehicle that shortens the critical paths by placing various macros and blocks on multiple layers [7]. At the architectural level, the improved bandwidth (and possibly improved capacity) has inspired new stacked cache structures [3]. The same trend has even inspired integrating/stacking memory [8] with the processor layer on a carrier. The amount of performance benefit achievable in such configurations is still heavily debated. In the case of die-to-wafer or wafer-to-wafer integration, the amount of on-stack variation can be quite significant, which may obscure the potential performance advantages. Also, these variations severely constrain the level of granularity at which functional partitioning can be done across layers. Similarly, in cache or memory stacking configurations, power delivery (i.e., C4 current limitations and thermal envelopes) are potential limiters.

*2.1.b. Modularity*
Though widely overlooked, modularity can be one of the main advantages that provide a potentially simpler design flow. To do this, large IP blocks (e.g., chiplets, or even layers - even in disparate technologies and from different vendors) can be incorporated into a stack. This requires that there be well-defined interfaces, communication protocols, and technology ground-rules that will be common to all of the individual components comprising the larger and more modular system. The overheads associated with such well-defined infrastructures, rules, and protocols are potentially lower than those required to compose the system using a traditional 2D approach. Without 3D, integrating disparate technologies into a single "socket" is sometimes impossible.

In this context, 3D integration becomes "an infrastructure" that brings the subsystems together by providing basic things such as interconnectivity, power delivery, and I/O. The corresponding modular systems can be rapidly customized for different customer/client needs and specifications without having to redesign the entire system from the ground up.

### 2.2. Challenges
The aforementioned potential leverages of 3D also bring in new risks and challenges. Both technology and system-level characterizations have been studied to try to understand those challenges more clearly [1], [4], [5]. By combining differently constrained layers into a stack, 3D integration will tend to impose the combined constraints on each individual layer. Among those constraints are: (i) a shared power envelope (with C4 current constraints), (ii) a shared thermal envelope for heat removal, and (iii) interactions between the layers (in the form of noise). While not a major problem in low power systems, the first two constraints can be quite challenging for high-power and high power-density applications, like microprocessors.

*2.2.a. Power Envelopes and Delivery*
The amount of current drawn in any layer can impact the other layers when there are shared Power/GND TSVs. Given that high-performance microprocessors are power-limited (and increasingly C4-current limited due to the power density issues), this can create major design constraints in a multi-layer 3D stack.

*2.2.b. Heat Removal and Reliability*
Thermal challenges have been a major limiter for many promising applications of 3D [10]. Increases in power density, along with increases in wiring density and wiring layers (which are required to wire more complex function, but which add thermal impedance) are real challenges. Even though low power memory chips (e.g.) have relatively low thermal profiles and few significant thermal challenges, a multitude of thermal effects need to be kept in mind when stacking them [9]. Not only do these affect the energy efficiency of the system, but they have important reliability implications as well.

Last but not least, the reduced distances in the vertical direction (especially in thinned silicon) will exacerbate reliability and noise issues. Stack-level analysis and optimization is essential in trying to address these concerns. This is a new area of research.

## 3. Beyond the Basics: Customized Modular Systems

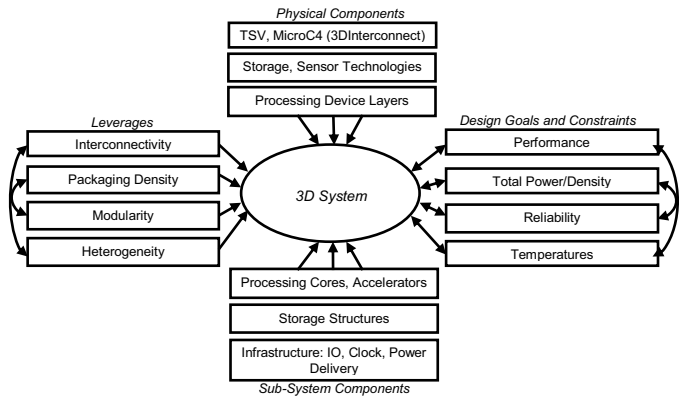

Figure 2. Composing a 3D system from sub-system components according to design goals and constraints

While Figure 2 illustrates the basic considerations, challenges, and opportunities for the design of a 3D system, it is far from being comprehensive. The basic limitation of optimizing for any single parameter (such as performance, or power, or density) has as obvious limitation: it is almost certain to degrade the other parameters in a 3D system.

Interactions among basic design goals and constrains are significantly more prominent and complex in 3D. Also, the additional complexity in each design stage adds to the overhead associated with iterations in other design stages.

For example, considering interconnectivity and packaging density together creates almost an ironic paradox. Specifically, increasing the number of TSVs to improve the interconnectivity between layers necessarily leads to a reduction in the active silicon area of each layer. As the area dedicated to TSVs and the corresponding keep-out-zones around them increase, the areal efficiency of each layer in the stack decreases.

As a different example, TSVs provide the inter-layer connectivity within a 3D stack. For systems having high levels of connectivity, this is most efficiently done by specifically allocating dense "via farms" on the chips. But the very structures that require high levels of connectivity tend also to have higher levels of computational activity–which are antithetical to the notion of "via farms." In addition, dense via farms (in which the vias are encased in electrical insulators) can also be thermal insulators, and will prevent heat from diffusing in planar dimensions. While benefitting the system by providing essential connectivity, they may insidiously cause a pernicious thermal problem.

## 4. Conclusions

The known values of 3D in the market today are clear: they enable better integration – they *simplify* the global system. Today, the obvious uses for 3D are the ones in which the costs, power, interconnectivity, and profit margins are all fairly low. How to practice 3D within these contexts is fairly well known, and indeed, already being practiced in many markets (cameras, DRAMs, cell phones, etc.).

What we've tried to do here is to articulate how 3D can be useful in more complicated systems that don't easily fathom these extensions today. The crux of doing this is (unsurprisingly) a "co-optimization" of both the system design and the system hardware across layers to facilitate modularity, flexibility, interconnectivity, and power/thermal efficiency.

3D offers some clear advantages in the future integration of systems: better volumetric density, lower raw power, smaller component count, and better modularity. But realizing these advantages requires solving a new set of problems in (literally) a new dimension.